\documentclass[longauth]{aa}
\usepackage{color}
\usepackage{graphicx}
\usepackage{epsfig}
\usepackage{enumerate}
\usepackage{amsmath}
\usepackage{booktabs}
\usepackage{mathtools}
\usepackage{float}
\usepackage{euscript}
\usepackage{chngcntr}
\usepackage[title,titletoc]{appendix}

\newcommand{\age}{$\tau$ }

\newcommand{\af}{[Mg/Fe] }

\newcommand{\vel}{km s$^{-1}$ }
\newcommand{\teff}{T$_\mathrm{eff}$ }
\newcommand{\rp}{R$_\mathrm{p}$ }
\begin{document}

\DeclareGraphicsExtensions{.ps,.pdf,.png,.jpg}
%\DeclareGraphicsRule{.ps}{pdf}{.pdf}{.jpg}
%\title{The Thick Thin Disk and the Thin Thick Disk: The Milky Way Has No Chemically Distinct Thick Disk}
%\title{The Thick Thin Disk and the Thin Thick Disk of the Milky Way with the AMBRE Project}
\title{ The AMBRE project: The thick thin disk and thin thick disk of the Milky Way}
%\title{The AMBRE Project: The Milky Way Has No Chemically Distinct Thick Disk}
%The Milky Way Has No Distinct Thick Disk

\author{M. R. Hayden\inst{1},
A. Recio-Blanco\inst{1},
P. de Laverny\inst{1},
S. Mikolaitis\inst{2},
C. C. Worley\inst{3}
}

\institute{%Hayden,Recio-Blanco, de Laverny, Guiglion
  Laboratoire Lagrange (UMR7293), Universit\'{e} de Nice Sophia Antipolis, CNRS, Observatoire de la C\^{o}te d'Azur, BP 4229, 06304 Nice Cedex 4, France  \label{inst1} (mhayden@oca.eu) 
  \and
  Institute of Theoretical Physics and Astronomy, Vilnius University, Saul\.{e}tekio al. 3, LT-10257, Vilnius, Lithuania \label{inst2}
  \and
  Institute of Astronomy, Cambridge University, Madingley Road, Cambridge CB3 0HA, UK  \label{inst3}
}
\titlerunning{The thick thin disk and thin thick disk}
\authorrunning{M. Hayden et al.}
\flushbottom
\abstract
{We analyze 494 main sequence turnoff and subgiant stars from the AMBRE:HARPS survey. These stars have accurate astrometric information from \textit{Gaia}/DR1, providing reliable age estimates with relative uncertainties of $\pm1$ or $2$ Gyr and allowing precise orbital determinations. The sample is split based on chemistry into a low-\af sequence, which are often identified as thin disk stellar populations, and a high-\af sequence, which are often associated with thick disk stellar populations. We find that the high-\af chemical sequence has extended star formation for several Gyr and is coeval with the oldest stars of the low-\af chemical sequence: both the low- and high-\af sequences were forming stars at the same time. We find that the high-\af stellar populations are only vertically extended for the oldest, most-metal poor and highest \af stars. When comparing vertical velocity dispersion for the low- and high-\af sequences, the high-\af sequence has lower vertical velocity dispersion than the low-\af sequence for stars of similar age. This means that identifying either group as thin or thick disk based on chemistry is misleading. The stars belonging to the high-\af sequence have perigalacticons that originate in the inner disk, while the perigalacticons of stars on the low-\af sequence are generally around the solar neighborhood. From the orbital properties of the stars, the high-\af and low-\af sequences are most likely a reflection of the chemical enrichment history of the inner and outer disk populations, respectively;  radial mixing causes both populations to be observed in situ at the solar position. Based on these results, we emphasize that it is important to be clear in defining what populations are being referenced when using the terms thin and thick disk, and that ideally the term thick disk should be reserved for purely geometric definitions to avoid confusion and be consistent with definitions in external galaxies.}

%From the orbital properties of the stars, it is likely that the high-\af and low-\af sequences are a reflection of the evolutionary history of the inner and outer disk populations respectively, rather than a thick and thin disk, with radial mixing causing both populations to be observed in situ at the solar position. Based on these results, we emphasize that it is important to be clear in defining what populations are being referenced when using terms like the thin and thick disk, and that ideally the term thick disk should be reserved for purely geometric definitions.}

%%%%%implying that the classical terminology of thin and thick disk for the chemical sequences identified in the solar neighborhood should be revised. It is likely that what we identify as the thick and thin disk are inner and outer disk populations (respectively) with radially mixing causing both populations to be observed insitu at the solar position. A more appropriate definition would be calling the high- and low-\af sequences as inner and outer disk, and reserving the term thick disk for purely geometric definitions.}

%\velspace{0.3cm}}
\maketitle
\section{Introduction} 
Unraveling the chemodynamic structure of the Milky Way is a key constraint for models of chemical and galaxy evolution. However, there has been much debate over the current structure of the disk (e.g., \citealt{Rix2013}), in which the interplay between the thin and thick disks is unclear. The advent of high-resolution spectroscopic surveys, along with data from the \textit{Gaia} \citep{Prusti2016,Brown2016} satellite, has the potential to unravel the current mysteries surrounding the formation and evolution of the Milky Way disk.

The thick disk was first identified by \citet{Yoshii1982,Gilmore1983} as an overdensity of stars at large distances from the Galactic plane. The stars of the thick disk in the solar neighborhood have been found to be $\alpha$-enhanced relative to the Sun (e.g., \citealt{Fuhrmann1998,Bensby2003,Adibekyan2013,Recio-Blanco2014}) and subsolar metallicity at around $-0.5$ dex. More recent observations have led to the discovery of intermediate $\alpha$ populations at much higher metallicities (e.g., \citealt{Bensby2007,Adibekyan2011,Hayden2015,Mikolaitis2017}), although there may be a prominent gap between the metal-poor high-\af populations and the more metal-rich intermediate-\af populations (e.g., \citealt{Adibekyan2011}). The origin of these intermediate-\af populations is difficult to discern. It is possible that they are chemically related to the high-\af metal-poor stars (e.g., \citealt{Haywood2016}), are a distinct population (e.g., \citealt{Jofre2017}), or could also result from extra mixing along the RGB for metal-rich populations \citep{Masseron2015}. If the high-\af metal-poor stars are chemically related with the intermediate-\af stars, there appears to be two distinct stellar population sequences in the solar neighborhood. One sequence starts at high-\af with the \af decreasing as [Fe/H] increases even past solar metallicities, and one sequence begins at solar-\af abundances spanning a wide range of metallicities from $\sim-0.6<\mathrm{[Fe/H]}<0.4$. The thick and thin disks are often identified in the \af versus [Fe/H] plane chemically; these intermediate-$\alpha$ high-metallicity stars are sometimes lumped in with the thick disk populations and the solar-\af populations are identified with the thin disk. However, the connections, if there are any, between the thin and thick disk chemical sequences is unclear. Furthermore, there is overlap between the sequences at the highest metallicities at solar-\af abundances and it is also uncertain to which sequence (or both) these stellar populations belong. It has been argued that what we identify as the thick disk in the chemical plane is not a distinct phase of the disk evolution, but is instead the chemical evolution track of stellar populations from the inner Galaxy that is brought to the solar neighborhood through radial mixing processes \citep{Haywood2013}. In this case, it is likely that the super solar-metallicity populations currently identified as thin disk are actually the metal-rich extension of the high-\af thick disk sequence. 

With the advent of precision astrometric measurements from \textit{Gaia}, we can now measure accurately ages, velocities, and orbital properties for large numbers of stars that can be used in conjunction with ground-based spectroscopic surveys. In this paper, we describe the AMBRE:HARPS dataset of main sequence turnoff (MSTO)/subgiant stars for which we obtain reliable ages and measure the chemical, kinematic, and temporal properties of the disk. We then discuss the structure of the disk in the solar neighborhood based on the HARPS-TGAS dataset. 

\section{Data}

\subsection{AMBRE:HARPS}
The AMBRE project is a uniform analysis of high-resolution archival ESO spectra, as described in \citet{deLaverny2013}. We focused on spectra taken from the HARPS instrument as this sample has large overlap with the \textit{Gaia}/TGAS catalog. Stellar parameters are derived using the MATISSE algorithm \citep{Recio-Blanco2006}, as described in \citet{DePascale2014} for the HARPS instrument. Individual abundances for this sample are derived in \citet{Mikolaitis2017}. In particular, our analysis uses the effective temperature measurements from AMBRE, along with the [Fe/H] and \af derived in the \citet{Mikolaitis2017} analysis. Typical errors are $<100$K in \teff and 0.05 dex in [Fe/H] and [Mg/Fe].
\subsection{\textit{Gaia} DR1}

Parallaxes and proper motions for the AMBRE:HARPS sample were taken from \textit{Gaia} DR1 \citep{Lindegren2016}. This sample is hereafter referred to as HARPS-TGAS and contains $\sim1 500$ stars. We computed distances using a Bayesian approach as outlined in \citet{Bailer-Jones2015}, consisting of a single disk with an exponential scale height of 300 pc and exponential scale length of 2.7 kpc. Additionally, we used the luminosity function derived by \citet{Robin2012Gaia}. As the majority of the sample is within 50 pc, the choice of prior has little impact on the overall results, as the fractional parallax error at these distances is very small ($<5$\%). 
\subsection{Ages}

We used Bayesian inference to determine ages for the HARPS-TGAS sample using methods similar to those described in \citep{Jorgensen2005} by comparing the derived stellar parameters to Dartmouth isochrones (\citealt{Dotter2008}; http://stellar.dartmouth.edu/models/isolf\_new.html). We select main sequence turnoff and subgiant stars for which the age estimates are most reliable. A more detailed description of the ages is given in Appendix A. The average ($1\sigma$) relative errors for the ages are 1-2 Gyr when using simulated observations from TRILEGAL \citet{Girardi2005} or comparing to the sample analyzed by \citet{Haywood2013}. The isochrones range in age from 0 to 15 Gyr, so the absolute age scale used in this analysis is somewhat stretched, particularly for the oldest stars.

\subsection{Velocities and orbits}
Velocities and orbits were determined via \textit{Galpy} \citep{Bovy2015a} using distances and proper motions from \textit{Gaia} DR1, along with radial velocities as determined by \textit{AMBRE}. For the orbit integration, we used the MWPotential2014 described in \citet{Bovy2015a} for the Galactic potential. We assume a solar position of 8 kpc and a rotational velocity of the disk of 220 \vel. Typical uncertainties in individual velocity components are ${\lesssim}1$ \vel and in the orbital parameters such as R$_\mathrm{p}$ or z$_\mathrm{max}$ of $\sim5$\%. Errors in the velocities and orbital parameters are computed using a Monte Carlo (MC) of the dervied and observed parameters (e.g., \teff, M$_\mathrm{K}$, $\mu$). The method for measuring the vertical velocity dispersion is described in \citet{Guiglion2015} and \citet{Hayden2017b}. We measured the vertical velocity dispersion in bins of 2 Gyr and required a minimum of eight stars per bin to compute the velocity measurements. 

\section{Results}
\begin{figure}[t!]
\centering
\includegraphics[trim=10bp 5bp 130bp 5bp,clip,width=3.3in]{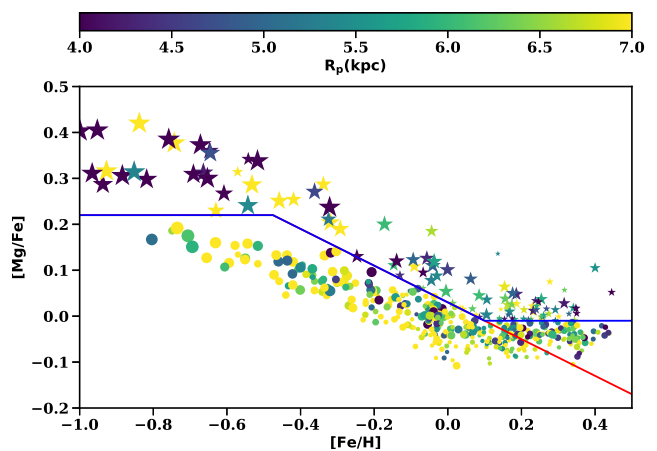}
\caption{ [Mg/Fe] vs. [Fe/H] plane for the HARPS sample with reliable age estimates. The blue line denotes our chemical separation between thin and thick disks, where the thick disk stars (star symbol) are shown above the line and thin disk stars (circle symbol) below the line. The size of each data point denotes the relative age of each star; the larger points are older than smaller points. Stars are color coded by their \rp. The color bar stops at \rp$=7$ kpc to prevent color saturation of stars coming from the inner Galaxy. Which sequence the most metal-rich, solar-\af stars belong to is unclear, and we show the red line as an example of a different potential split between stellar population groups. }
\label{mgfe}
\end{figure}
\subsection{Ages of the thin and thick disk stars}

\begin{figure}[t!]
\centering
\includegraphics[trim=10bp 10bp 20bp 0,clip,width=3.3in]{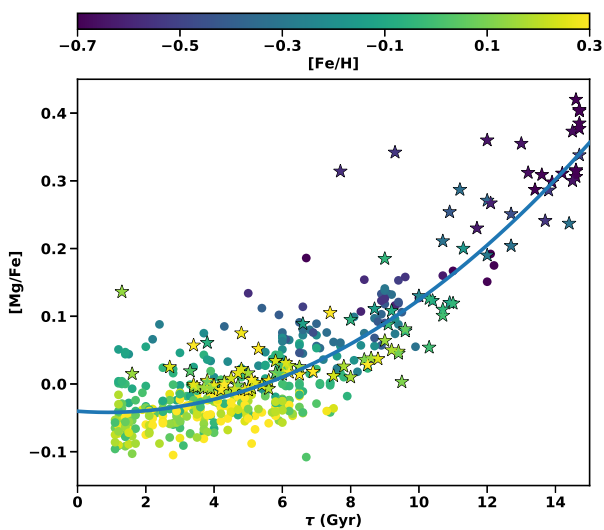}
\caption{[Mg/Fe] vs. age relation for the HARPS sample, with stars color-coded by metallicity. The symbols used are the same as in Fig. \ref{mgfe}, where the high-\af track is represented by stars and the low-\af track as circles.}
\label{agealpha}
\end{figure}

We separated the stars into low- and high-\af sequences based on a combination of their \af and [Fe/H] abundances as shown by the blue line in Fig. \ref{mgfe}; the high-\af stars (star symbols) represent those above the blue line, while the low-\af stars (circles) represent those below the blue line. Individual stars are color coded by the estimated perigalacticon R$_\mathrm{p}$. The \rp is the closest approach to galactic center of the orbit of a star and is a useful tracer, particularly for the most metal-rich populations, in constraining the relative importance of the different mechanisms responsible for radial mixing of stellar populations (i.e., blurring versus churning; see \citealt{Sellwood2002}). The color bar has a cutoff of \rp$=7$ kpc, but many of the stars actually have larger \rp that extend all the way to the solar radius at $\sim8$ kpc. The cutoff prevents color saturation of the stars belonging to the inner disk. 

\begin{figure}[t!]
\centering
\includegraphics[trim=35bp 10bp 0 55bp,clip,width=3.3in]{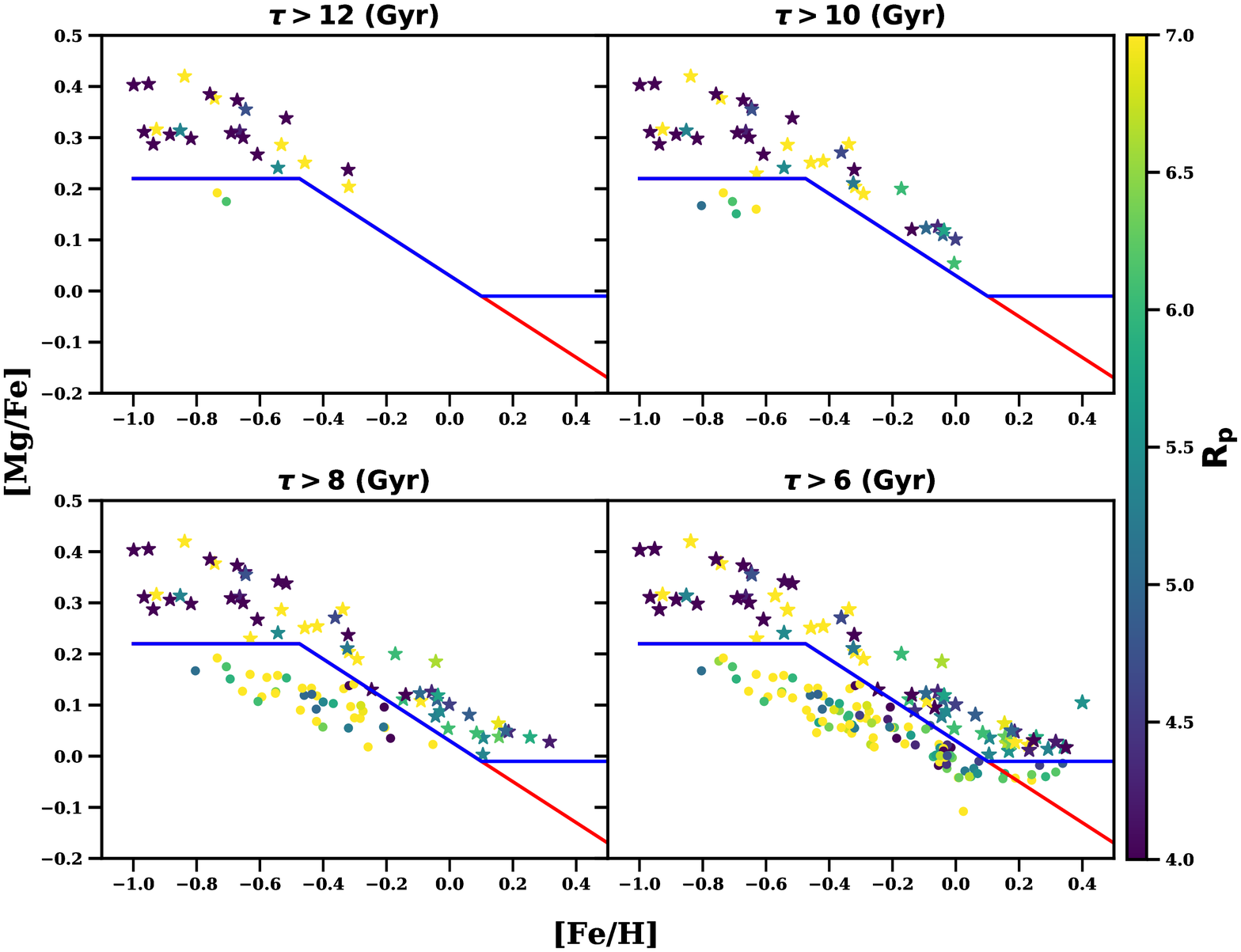}
\caption{[Mg/Fe] vs. [Fe/H] plane as a function of various age ranges, going from oldest to younger stellar populations. Stars are color coded by their perigalacticons R$_\mathrm{p}$. The blue and red lines are the same as in Fig. \ref{mgfe}, where the high-\af population are denoted by star symbols and the low-\af population are denoted by circles. }
\label{agerange}
\end{figure}

In Fig. \ref{agealpha} we find that the oldest stellar populations, with \age$>12$ Gyr, belong to the high-\af sequence (star symbols) and are the most metal-poor and high-$\alpha$ stars in the sample. Many of these stars come from the inner Galaxy, with \rp$<5$ kpc (Fig. \ref{mgfe}), and are likely associated with the classical thick disk. We find that there is extended star formation for many Gyr along the high-\af sequence, as shown in Fig. \ref{agerange}. Following the high-\af track, we find that the \af ratio begins to decrease at [Fe/H]$\sim-0.5$, as Type Ia supernovae begin to pollute the ISM. As shown in Fig. \ref{mgfe} and \ref{agealpha} there is a clear age-$\alpha$-metallicity relation for the thick disk stars up until $\tau\sim6$ Gyr ([Fe/H]$\sim0.2$), where the \af ratio decreases and metallicity increases for younger stellar populations. For stars along the high-\af sequence, we see a general trend in increasing \rp with increasing metallicity: that is to say, the orbits for younger stars along the high-\af sequence are more circular than those of older high-\af stars. It is unclear to which stellar populations the most metal-rich solar-\af stars belong: it is possible that the stars with [Fe/H]$>0.2$ that are currently labeled as thin disk are in fact the most metal-rich extension of the high-\af chemical track. As an example, the red line in Fig. \ref{mgfe} shows a different potential split between stellar populations for which the highest metallicity populations all belong to the high-\af sequence. In this case, star formation along the thick disk sequence has continued up to present day. 

Along the low-\af track, we find that the oldest stars appear $\sim10$ Gyr ago at intermediate \af abundances and low ($<-0.6$ dex) metallicity. We define a star as having a roughly circular orbit if it has $e<0.2$. We find that many of these older thin disk stars are on circular orbits in the solar neighborhood or come from larger radii with apogalacticons greater than 10 kpc; very few of such thin disk stars are on orbits that stray into the inner disk. The thin disk sequence has a slight \age-$\alpha$-metallicity relation for older, more metal-poor populations. As age increases, $\alpha$ decreases and metallicity decreases. However, this relationship breaks down at around $\sim6$ Gyr, as stars with these ages have a large range in metallicity. For \age$<6$ Gyr, there is a spread for $-0.5<\textrm{[Fe/H]}<0.4$ in metallicity: for these stars there is no age-$\alpha$-metallicity relation and there is a large dispersion in the derived metallicities for the youngest stellar populations. For the most metal-rich stars on the thin disk track ([Fe/H]$>0.1-0.2$), we find that more than half are on circular orbits ($e<0.2$), despite their likely origin in the inner Galaxy based on their high metallicities. 

Finally, we find significant age overlap between the low- and high-\af sequences for many Gyr between $\sim6<\tau<10$ Gyr: the stellar populations belonging to the low- and high-\af populations are coeval and both tracks formed stars of very different metallicities at the same time, as shown in Fig. \ref{agerange}.% We note again that the most metal-rich stars could belong to either sequence (or both), and the separation between thin and thick disk in the chemical plane at the metal-rich end is unclear and our blue line somewhat arbitrary, as the red line (or a different split between stellar populations) is equally justifiable. 
 
\subsection{Vertical distribution of the low- and high-[Mg/Fe] tracks}

\begin{figure}[t!]
\centering
\includegraphics[width=3.5in]{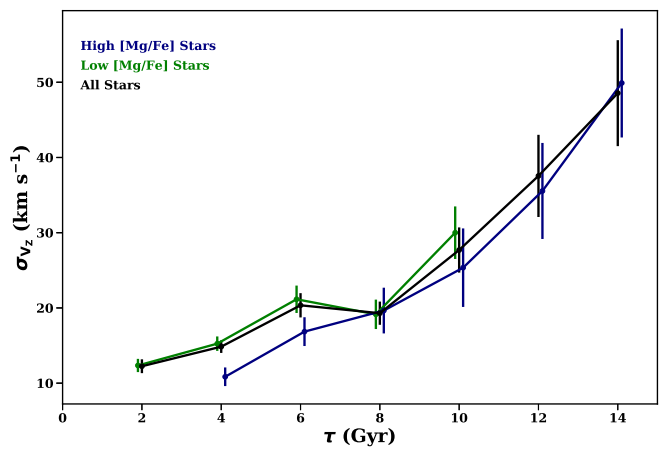}
\caption{Vertical velocity dispersion as a function of age for the low-\af (green), high-\af (blue), and total (black) stellar sample. Velocity dispersion is measured in bins of 2 Gyr, where each bin has a minimum of 8 stars per bin.}
\label{ttdisk}
\end{figure}

%We measure the \zm- which is the maximum height above the plane a star reaches during it's orbit- for individual stars and compute the median \zm for both the low- and high-\af sequences as a function of age, as shown in the left panel of Fig. \ref{ttdisk}. We find that the high-\af populations (blue line) are vertically extended only for the oldest, highest-$\alpha$ populations, with the vertical extent of the high-\af stars dropping rapidly with age. The behavior of the solar-\af populations (green line) is quite similar to that of the high-\af as a function of age, with the median \zm gradually increasing as age increases. Not only are the global trends similar, we find that the low-\af populations are just as vertically extended, if not more so, than the high-\af populations at the same age, in particular where there is significant overlap of stellar populations ($6<\tau<10$ Gyr)!
%This is also clearly seen in the vertical velocity dispersion for the different stellar populations in the right panel of Fig. \ref{ttdisk}. 
Using bins of 2 Gyr, we measured the vertical velocity dispersion for both the low- (green line) and high-\af (blue line) sequences as shown in Fig. \ref{ttdisk}. We find little evidence for a step function in the vertical velocity dispersion, and instead find that the vertical velocity dispersion generally smoothly increases with age for both tracks, save for a dip at $\tau\sim8$ Gyr for the older thin disk stars. This means that the high-\af sequence is only vertically extended for the oldest, highest \af populations and the vertical extent of the high-\af stars drops rapidly with age. Most striking is that the velocity dispersion for the high-\af track is, at the same age, always equal to or lower than the low-\af track. The high-\af sequence is no more vertically extended than the low-\af sequence when the populations are coeval.

\section{Discussion}
%This means that the chemical thick disk is not a completely separate population from the rest of the disk
We find that the high-\af sequence has an extended star formation history, potentially up to the present day and that the low- and high-\af sequences have significant age overlap and are coeval. Chemically, there are two distinct sequences forming stars at the same time at vastly different metallicities. The high-\af sequence is traditionally identified with the thick disk, but is only vertically extended for the oldest, highest \af, most metal-poor populations. At the ages where the low- and high-\af sequences overlap, there is little difference in the vertical heights of the two populations.

We also note that our results could be dependent on the selection function for the AMBRE:HARPS sample. However, these results have been observed previously by \citet{Bovy2012b} and \citet{Bovy2016}, using SEGUE and APOGEE data, respectively, but were not commented on extensively in these papers. Additionally, we note that any large errors in age determination could cause an error in the measurement of the velocity dispersion. 

What we are likely seeing is the mixing of two stellar populations coming from different locations:  the high-\af sequence, which formed in the inner Galaxy where star formation rates were significantly higher than the low-\af sequence, which formed stars at much lower rates and had a more gradual build up of metallicity. Stars belonging to the high-\af sequence are observed in the solar neighborhood via radial mixing, wherein the older stars are primarily blurred to the solar cylinder, while those of higher metallicity and lower \af come from both blurring and churning mechanisms. The idea that the solar neighborhood is the juxtaposition of the inner and outer disk is not new. \citet{Haywood2013} argued that this is a potential explanation for the observations of the two sequences in the solar neighborhood, and the potential mixing of the high-\af sequence with a more in situ population was initially seen in Fig. 20 of \citet{Edvardsson1993}, who found that stars coming from the inner disk were often metal poor and \af enhanced, while more in situ stars generally had solar type abundances. Qualitatively, our results are very similar to those of \citet{Edvardsson1993} with respect to the inner disk being dominated by high-\af populations. With the addition of extremely precise kinematics and distances from \textit{Gaia} and a larger sample size of stars with reliable ages, we find that the high-\af sequence was likely not a completely separate phase of disk evolution but indeed formed stars for many Gyr and was coeval with the the low-\af sequence of populations. This inner disk origin for the bulk of the high-\af sequence explains the relative similarity of the high-\af sequence with radius (e.g., \citealt{Nidever2014}), as radially mixing of the high-\af population out to larger radii preserves its appearance in the chemical plane. 

Turning our attention to the super solar-metallicity populations, many of these stars are on circular orbits and never reach the radii at which they likely formed based on their high metallicity and the observed radial abundance gradients in the disk (e.g., \citealt{Hayden2014}). For these populations, churning is an important mechanism for bringing them to the solar neighborhood, which is in agreement with observations from \citet{Kordopatis2015} and \citet{Hayden2017b}. If we consider the super-solar metallicity stars as the metal-rich extension of the high-\af sequence, the relative importance of various mixing processes for the high-\af sequence depends on the age/metallicity of the stellar populations: older (more metal-poor) populations are most likely observed in the solar neighborhood via blurring, while the younger (more metal-rich) populations appear to be radially mixed via churning.

Based on these arguments, we find that the current chemical definitions of thin and thick disk are misleading. We urge that the term thick disk be reserved for the geometric thick disk, with the realization that the chemistry of stars making up the geometric thick disk vary depending on the radius at which they are observed. In the inner Galaxy and solar neighborhood, the geometric thick disk is dominated by metal-poor high-\af populations, however in the outer disk the geometric thick disk is dominated by flaring solar-\af populations (e.g., \citealt{Hayden2015,Minchev2015,Bovy2016,Mackereth2017}). Chemically, the two sequences observed in the \af plane in the solar neighborhood are mostly likely a reflection of their birth radius, rather than a thin or thick disk, and the sequences are coeval. Inner and outer disk stellar populations is a less ambiguous definition for these sequences, as we have shown that calling the high-\af sequence the chemical thick disk is a misleading and these populations are no thicker than the chemical thin disk for stars of the same age.

\begin{acknowledgements}
The authors thank Misha Haywood for providing derived ages from their 2013 paper and for useful discussions. MRH and ARB received financial support from ANR, reference 14-CE33-0014-01. This work has made use of data from the European Space Agency (ESA) mission {\it Gaia} (\url{https://www.cosmos.esa.int/gaia}), processed by the {\it Gaia} Data Processing and Analysis Consortium (DPAC; \url{https://www.cosmos.esa.int/web/gaia/dpac/consortium}). Funding for the DPAC has been provided by national institutions, in particular the institutions participating in the {\it Gaia} Multilateral Agreement. 
\end{acknowledgements}

\bibliographystyle{aa}
\bibliography{ref}

\begin{appendices}
\counterwithin{figure}{section}
\section{Ages} 
As noted above, we used Bayesian inference to determine ages to individual stars by comparing derived stellar parameters to isochrones. The Dartmouth isochrones \citep{Dotter2008} used have metallicities ranging from $-2.0<\mathrm{[Fe/H]}<0.6$ in steps of 0.1 dex, are linearly spaced in age in steps of 100 Myr from 0.1 Gyr to 15 Gyr, and have \af spacing of 0.2 dex from $0.0<\mathrm{[Mg/Fe]}<0.4$. The isochrones and the AMBRE stellar parameters are on a slightly different temperature scale based on the V-K colors of the stars. We find an offset of 100K using the color-temperature relations derived in \citet{Gonzalez-Hernandez2009} between the temperatures derived from AMBRE and those of the isochrones, and apply a shift of 100K to the \teff to place the derived AMBRE temperatures on the same scale as the isochrone temperatures. A complete description of the method, along with a catalog of the ages presented in this paper, will be described in a later paper \citep{Hayden2017a}. 

\begin{figure}
\label{haycomp}
\includegraphics[width=3.5in]{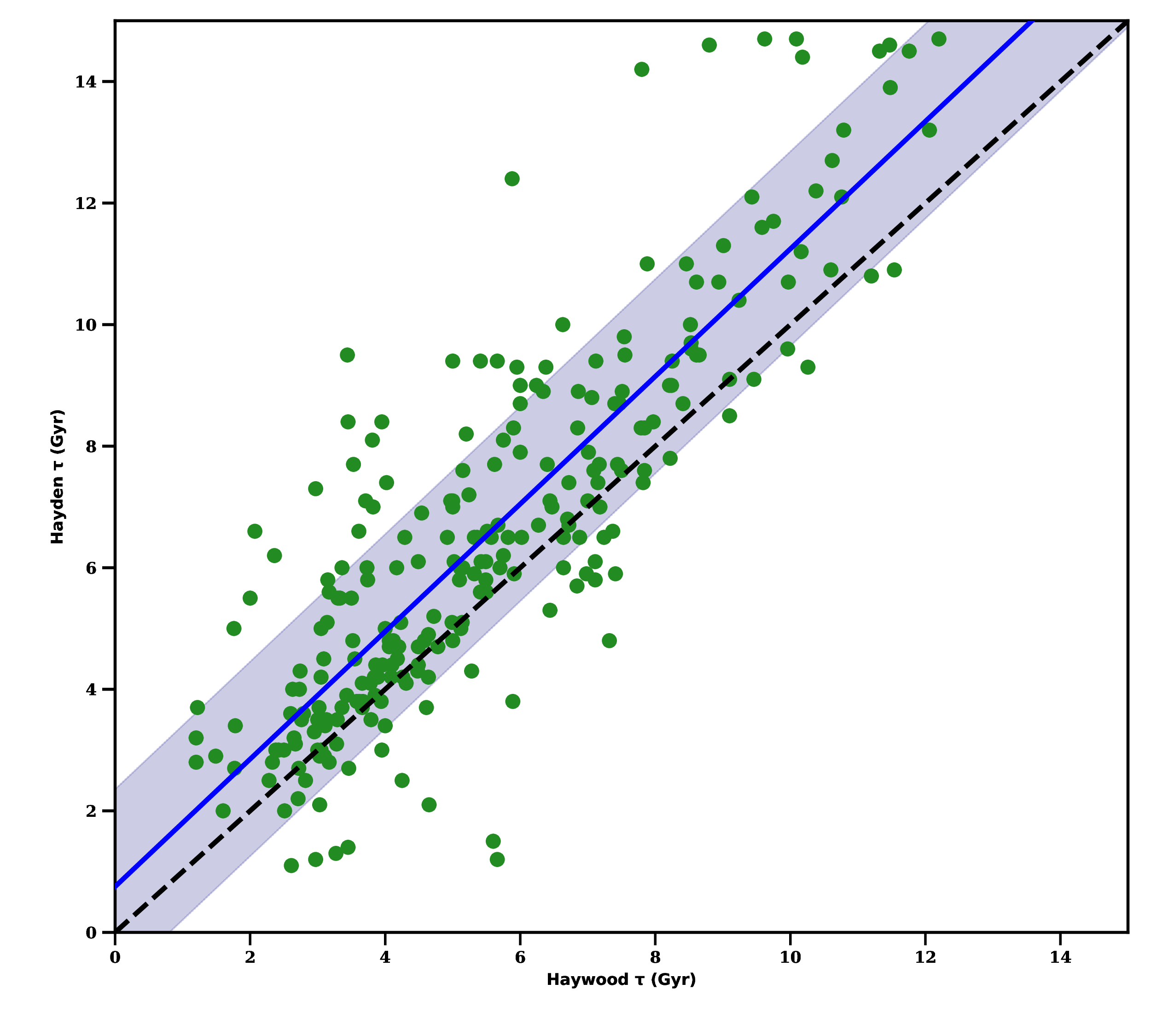}
\caption{Comparison between our derived ages and those provided by \citet{Haywood2013}. The dashed black line denotes the 1:1 relation, while the blue line denotes the relation between our derived ages and those of Haywood with a slope of 1.03. The blue shaded region shows the $1\sigma$ scatter of 1.6 Gyr.}
\end{figure}

To determine the stellar ages, we adopt several simple priors. We assume a uniform star formation history and additionally weight each isochrone point by mass using a Chabrier IMF \citep{Chabrier2003}. No prior for age as a function of metallicity or \af is used (i.e., we do not force a high-\af or metal-poor star to be old). The mode of the generated PDF is used to characterize the age of an individual star. While carrying out tests with simulated observations using stars from TRILEGAL, we find ages are reliable (relative errors $<1-2$ Gyr) for turnoff and subgiant stars only, which are selected by requiring $\mathrm{M_J}<3.75$ and $3.6<\log{\mathrm{g}}<4.4$. From our initial sample of nearly 1,500 stars, 494 meet our selection criteria. Additionally, we compared these remaining stars to stars in common with \citep{Haywood2013} as shown in Fig. \ref{haycomp}. We find a systematic offset of 1 Gyr, identical to the right panel of Fig. 3 of \citet{Haywood2013} when comparing their method using Dartmouth and Y$^2$ isochrones. Other than the systematic offset, the ages are very similar, with a linear 1:1 relation found between the age estimates and a random scatter of 1.6 Gyr between the samples, despite using independent stellar parameters, abundances, and isochrone sets. The largest outliers are driven primarily by differences in temperature between AMBRE and those adopted by \citet{Haywood2013}. Additionally, for the oldest stars the different age ranges covered by the isochrones also has an impact.

%\subsection{Velocities and Orbital Parameters}
\end{appendices}

\end{document}